\documentstyle[12pt,bezier]{article}

\newcommand{\be}{\nopagebreak\begin{equation}}
\newcommand{\ee}{\end{equation}}
\newcommand{\ba}{\begin{array}}
\newcommand{\ea}{\end{array}}
\newcommand{\bp}{\begin{picture}}
\newcommand{\ep}{\end{picture}}

\newcommand{\rf}[1]{Ref.~\cite{#1}}
\newcommand{\eq}[1]{Eq.~(\ref{#1})}
\newcommand{\bi}[6]{\bibitem{#1}{#2, }{\sl #3 }{\bf #4}{ (#5)}{ #6}}
\newcommand{\wt}[1]{\widetilde{#1}}
\newcommand{\wh}[1]{\widehat{#1}}
\newcommand{\ie}{{\em i.e., }}

\renewcommand{\d}{\partial}
\newcommand{\eps}{\varepsilon}
\newcommand{\ph}{\varphi}
\newcommand{\ps}{\psi}
\newcommand{\tr}{{\rm tr}\,}
\newcommand{\scr}{\scriptstyle}
\newcommand{\scs}{\scriptscriptstyle}
\newcommand{\R}{{\scriptscriptstyle R}}
\newcommand{\F}{{\scriptscriptstyle F}}
\newcommand{\N}{{\scriptscriptstyle N}}

\renewcommand{\P}{{\cal P}}

\begin{document}
\begin{titlepage}
\begin{flushright}
SISSA-24/95/EP\\
March, 1995
\end{flushright}
\vspace*{36pt}
\begin{center}
{\Huge \bf One-dimensional dynamics of\\ QCD$_2$ string}
\end{center}
\vspace{2pc}
\begin{center}
 {\Large D.V. Boulatov}\\
\vspace{1pc}
{\em International School for Advanced Studies (SISSA/ISAS)\\
via Beirut 2-4, I-34013 Trieste, Italy}
\vspace{2pc}
\end{center}
\begin{center}
{\large\bf Abstract}
\end{center}
We show that QCD$_2$ on 2D pseudo-manifolds is consistent with the
Gross-Taylor string picture. It allows us to introduce a model
describing the one-dimensional evolution of the QCD$_2$ string (in the
sense that QCD$_2$ itself is regarded as a zero-dimensional system).
The model is shown to possess the third order phase transition
associated with the $c=1$ Bose string below which it becomes
equivalent to the vortex-free sector of the 1-dimensional matrix
model. We argue that it could serve as a toy model for the
glueball-threshold behavior of multicolor QCD.
\vfill
\end{titlepage}

\section{Introduction}

QCD on 2-dimensional compact manifolds attracted attention for the
first time in \rf{R1} and then was investigated from different
viewpoints in Refs.~[2-9] (the earlier papers not touching the
topological aspects are \cite{M,GW,KK}).  The recent renewal of
interest in 2-dimensional gauge theories was in a big part triggered
by the Gross and Taylor stringy picture of QCD$_2$ \cite{GT}. It
appears that, in the large $N$ limit, the spherical topology is
distinguished from all the others \cite{R2}. In this case, continuous
QCD$_2$ undergoes the third order phase transition below which the
model apparently admits of no stringy interpretation \cite{DK}. This
transition also takes place on all simply-connected closed
2-dimensional pseudo-manifolds, the so-called homotopic bouquets of
spheres. The simplest example of such a space is given by $p$ disks
whose boundaries are identified. The Euler character of this object
equals $p$. Two disks give a sphere.  If $p\geq3$, we obtain the
simply-connected pseudo-manifold $\P_p$.

The sum-over-coverings picture has limited validity for 2D spaces
having a non-trivial second homotopy group, $\pi_2$. Let us consider
the Wilson average for a simple closed loop: $W(L)$. If the loop
shrinks, $L\to\bullet$, then $W(\bullet)=1$ in any gauge theory, while
within a string model one finds a closed-string partition function
with one puncture.  As $\pi_2\neq0$, there is no geometrical reason
for the string partition function to vanish \cite{Wloop}. Since all
plaquette-made lattices can be regarded as 2D spaces with non-trival
$\pi_2$, this situation is not exotic.

As was discovered and Douglas and Kazakov \cite{DK}, the nice stringy
picture is spoiled when the manifold is a 2-sphere of a small enough
area.  Although it allowed for some speculations, the singleness of
this example restricts very much our intuition with respect to
possible guesses about more realistic physical systems. Fortunately,
the consideration can be extended in two directions without loosing
exact solubility. First, one can consider QCD$_2$ on the
pseudo-manifolds ${\P_p}$.  And second, one can make the model
dynamical by introducing a continuous time direction along which the
QCD$_2$ string can propagate.  Let us imagine a string theory in which
no internal degrees of freedom can be exited. The only allowed
processes are creation, destruction, splitting and joining of closed
strings. As will be discussed later, such a toy model can be quite
instructive and even help us to learn something about more realistic
physical systems.

\section{QCD$_2$ on pseudo-manifolds}

According to the general rules \cite{W,M}, we can construct the
QCD$_2$ partition function on $\P_p$ by putting into correspondence to
a disk of a (dimensionless) area $A$ the $U(N)$ heat-kernel

\be
G_A(\Omega)=\sum_R d_\R e^{-\frac{A}{2N}C_R}\chi_\R(\Omega)
\label{disk}
\ee
where $\Omega\in U(N)$ is a holonomy along the disk boundary. We
obtain the partition function on $\P_p$ by identifying the holonomies
and integrating over them:

\be
Z_{\P_p}(A_1,\ldots,A_p)=\int d\Omega \prod_{k=1}^p G_{A_k}(\Omega)
=\sum_{R_1\ldots R_p}
\prod_{k=1}^p\bigg(d_{\R_k}e^{-\frac{A_k}{2N}C_{R_k}}\bigg)
\int d\Omega \prod_{k=1}^p \chi_{\R_k}(\Omega)
\label{p-disks}
\ee

In Eqs.~(\ref{disk}) and (\ref{p-disks}), $R_1,\ldots,R_p$ are $U(N)$
irreps parametrized by lengths of rows in Young tables:

\be
R\equiv[m_1,m_2,\ldots,m_\N] \hspace{2pc}m_1\geq m_2\geq\ldots\geq m_\N
\label{hwc}
\ee
Technically, it is more convenient to introduce the strictly ordered
numbers

\be
\ell_k=m_k+\frac{N+1-2k}2, \hspace{2pc} \ell_1>\ell_2>\ldots>\ell_\N
\ee
in terms of which the dimension of an irrep $R$ takes the form

\be
d_\R=\prod_{i<j}\bigg(\frac{\ell_i-\ell_j}{j-i}\bigg)=
\frac{\Delta(\ell)}{\Delta_0}
\ee
$\Delta(\ell)$ is the Van-der-Monde determinant;
$\Delta_0=\Delta(\ell)|_{\ell_k=k}$.
The second Casimir eigenvalue is

\be
C_R=\sum_{k=1}^N \ell_k^2-\frac{N(N^2-1)}{12}
\ee
The first Weyl formula represents the character as the ratio of the
two determinants:

\be
\chi_\R(e^{i\ph})=\frac{\det (e^{i\ph_j\ell_k})}
{\det(e^{i\ph_j(\frac{N+1}2-k)})}
\ee

The integral of $p$ characters gives the multiplicity of the
trivial representation in the tensor product
$R_1\otimes R_2\otimes\ldots\otimes R_p$. This non-trivial factor is
characteristic to the pseudo-manifolds in question. We want to show
that its presence does not destroy the Gross-Taylor sum-over-coverings
picture of QCD$_2$. More precisely, we are going to show that, once
one interprets the number of boxes in a Young table as the number of
sheets of a covering, the multiplicities allow only for the tables
compatible with the geometrical interpretation. It is important that
the coverings are oriented. The change of an orientation of a disk
corresponds to the conjugation of a representation attached to it:

\be
\Big(R=[m_1,m_2,\ldots,m_{\scs{N-1}},m_\N]\Big)\Rightarrow
\Big(\wt{R}=[-m_\N,-m_{\scs{N-1}},\ldots,-m_2,-m_1]\Big)
\ee

The associativity of the tensor product means that it is sufficient to
consider only the three irrep multiplicities

\be
M_{R_1R_2}^{R_3}=\int_0^{2\pi} \prod_{n=1}^N d\ph_n |\Delta(e^{i\ph})|^2
\chi_{\scs{R_1}}(e^{i\ph})\chi_{\scs{R_2}}(e^{i\ph})
\overline{\chi_{\scs{R_3}}(e^{i\ph})}
\ee
Geometricly, it means that we can always deform a pseudo-manifold in
such a way that only 3 disks meat on every boundary circle. One can
imagine the corresponding construction as a number of cylinders glued
pairwise along boundaries of disks (see an example in Figure~1). If
the cylinders shrink to circles, the described above construction
recovers.

\begin{figure}
\begin{center}
\setlength{\unitlength}{1.00mm}
\linethickness{0.6pt}
\begin{picture}(70,30)(0,-15)
\bezier{150}(5,-10)(1,0)(5,10)
\bezier{150}(5,-10)(9,0)(5,10)
\bezier{150}(15,-10)(11,0)(15,10)
\bezier{150}(15,-10)(19,0)(15,10)
\bezier{20}(25,-10)(21,0)(25,10)
\bezier{150}(25,-10)(29,0)(25,10)
\bezier{150}(35,-10)(31,0)(35,10)
\bezier{150}(35,-10)(39,0)(35,10)
\bezier{150}(45,-10)(41,0)(45,10)
\bezier{150}(45,-10)(49,0)(45,10)
\bezier{20}(55,-10)(51,0)(55,10)
\bezier{150}(55,-10)(59,0)(55,10)
\bezier{150}(65,-10)(61,0)(65,10)
\bezier{150}(65,-10)(69,0)(65,10)
\put(15,-10){\line(1,0){10}}
\put(15,10){\line(1,0){10}}
\put(45,-10){\line(1,0){10}}
\put(45,10){\line(1,0){10}}
\end{picture}
\raisebox{1.5cm}{$\Longrightarrow$}
\begin{picture}(30,30)(0,-15)
\bezier{150}(5,-10)(1,0)(5,10)
\bezier{150}(5,-10)(9,0)(5,10)
\bezier{20}(15,-10)(11,0)(15,10)
\bezier{150}(15,-10)(19,0)(15,10)
\bezier{20}(25,-10)(21,0)(25,10)
\bezier{150}(25,-10)(29,0)(25,10)
\put(5,-10){\line(1,0){20}}
\put(5,10){\line(1,0){20}}
\end{picture}
\caption[x]
{\hspace{2cm}\parbox[t]{10cm}
{\small An example of a pseudo-manifold: $\wh{\P}_3$.}}
\label{Fig.1}
\end{center}
\end{figure}

The QCD$_2$ partition function on a cylinder of an area $\eps$ with
fixed holonomies along its boundary loops is

\be
Z^{(1)}_{\eps}(\Omega_1,\Omega_2)=
\sum_R e^{-\frac{\scr{\eps}}{2N}C_R}\chi_\R(\Omega_1)
\overline{\chi_\R(\Omega_2)}
\label{cylinder}
\ee
Thus we find for the deformed pseudo-manifold $\wh{\P_p}$

\be
Z_{\wh{\P}_p}=\sum_{R_1\ldots R_p}\sum_{S_1\ldots S_{p-1}}
\prod_{k=1}^p\bigg(d_{\R_k}e^{-\frac{A_k}{2N}C_{R_k}}
e^{-\frac{\scr{\eps}}{2N}C_{S_k}}\bigg)
\prod_{k=1}^p M^{S_k}_{R_kS_{k-1}}
\label{pipe}
\ee
where it is assumed that $S_0\equiv 0$ and $S_p\equiv0$ (hence
$S_1=R_1$ and $S_{p-1}=\wt{R_p}$).

We would like to treat coverings having opposite orientations
separately. It is self-consistent only in the $N\to\infty$ limit and
only if one restricts the consideration to irreps for which the second
Casimir is of the maximum order $N$ \cite{GT}. Then the theory
possesses 2 chiral sectors:\\
1)~Irreps having a finite number of positive highest weight components

\be
\{R_+\}:\ (m_1\geq m_2\geq\ldots\geq m_n\geq m_{n+1}=m_{n+2}=\ldots=m_\N)
\ee
We shall associate irreps of this form with coverings of the positive
orientation.\\
2)~Irreps having a finite number of negative highest weight components

\be
\{R_-\}:\ (0=m_1=\ldots=m_{n-1}\geq m_n\geq\ldots\geq m_\N)
\ee
are associated with the inverse orientation. The trivial
representation, $\{0\}$, corresponds to the empty configuration.

The general case is described by a product of 2 irreps from the
different sectors:

\be
R=r_+\otimes r_-,\hspace{2pc} r_+\in\{R_+\}\hspace{1pc}
r_-\in\{R_-\}
\ee
A decomposition of $R$ into irreducible representations is determined
by $r_+$ and $r_-$, and in turn, uniquely determines them provided
the mentioned above conditions are fulfilled.

Associativity of the tensor product again allows us to restrict the
consideration to irreps from only one of the sectors, because the
change of an orientation permutes $\{R_+\}$ and $\{R_-\}$. Let us
consider the multiplicity $M^T_{RS}$, where $R,\ S,\ T\in\{R_+\}$. The
corresponding pseudo-manifold consists of 3 cylinders glued along a
circle. We fixe 3 holonomies along the 3 components of
the boundary. To remove the branching points, we put into
correspondence to every cylinder the weight (cf. \eq{cylinder})

\be
H(u,w)=\sum_{r\in\{R_+\}}\chi_r(u)\overline{\chi_r(w)}=
\prod_{i,j=1}^N\frac1{1-u_i\overline{w_j}}=e^{F(u,w)}
\ee
where

\be
F(u,w)=\sum_{p=1}^{\infty}\frac1p\tr(u^p)\overline{\tr(w^p)}
\ee
is the generating function for $p$-fold unbranched connected coverings
of a cylinder. Being exponentiated, it produces all possible coverings
with equal weights and correct symmetry factors. It should be noted
that these simple expression and clear interpretation exist only for
coverings of a fixed orientation.

We have

\be
H(u_1,w)H(u_2,w)=\sum_{R,S,T\in\{R_+\}}M^T_{RS}\chi_\R(u_1)\chi_{\scs{S}}(u_2)
\overline{\chi_{\scs{T}}(w)}
\ee
As $\log \Big[H(u_1,w)H(u_2,w)]=F(u_1,w)+F(u_2,w)$, the appearing
configurations are exactly all possible oriented coverings of the 3
glued cylinders. Thus we arrive at the desired interpretation of
the multiplicities.

\section{Continuum limit in the infinite chain}

Let us consider the infinite pseudo-manifold of the type shown in
Fig.~1, \ie the infinite in both directions chain of cylinders and
disks.  According to the results of the previous section, the QCD$_2$
partition function for this object allows for the string
interpretation. One can regard this model as the ordinary lattice QCD
on the 1-dimensional lattice of cubes. We are looking for a continuum
limit in this system.

The partition function for each cylinder in the chain is given in
\eq{cylinder}. The sum over $U(N)$ irreps can be calculated
explicitely:

\[
Z^{(1)}_{\eps}(e^{i\ph},e^{i\ps}) = \sum_{\ell_1>\ldots>\ell_N}
e^{-\frac{\scr{\eps}}{2N}(\sum\ell_i^2-\frac{N(N^2-1)}{12})}
\frac{\det(e^{i\ph\ell})}{\Delta(e^{i\ph})}
\frac{\det(e^{-i\ps\ell})}{\Delta(e^{-i\ps})}
\]\[
=\frac{e^{\frac{\scr{\eps}(N^2-1)}{24}}}{\Delta(e^{i\ph})\Delta(e^{-i\ps})}
\frac1{N!}\sum_{\{\ell\in Z\}}\sum_{\P_1\P_2}(-1)^{\P_1\P_2}
e^{-\frac{\scr{\eps}}{2N}\sum\ell_k^2
+i\sum\ell_k(\ph_{\mbox{}_{\P_1k}}-\ps_{\mbox{}_{\P_2k}})}=
\]\be
=\frac{e^{\frac{\scr{\eps}(N^2-1)}{24}}}{\Delta(e^{i\ph})\Delta(e^{-i\ps})}
\sum_{\{h\in Z\}}\sum_{\P}(-1)^{\P}\bigg(\frac{N}{\eps}\bigg)^{\frac{N}2}
e^{-\frac{N}{2\scr{\eps}}\sum(\ph_{\mbox{}_k}+
2\pi h_k-\ps_{\mbox{}_{\P k}})^2}
\ee
where we have used the Poisson resummation formula. In the continuum
limit, the areas of the cylinders, $\eps$, tend to 0: $\eps\ll1/N$.
Thus we find the $N$ fermion kinetic term in the path integral. We can
take into account the winding numbers $h_k$ by considering the angles
$\ph_k$ and $\ps_k$ as unrestricted continuous variables.
It is well known that QCD$_2$ on a cylinder is equivalent to free
fermions \cite{D}.

For each disk we have the heat-kernel, which we have to expand up to
the first order in $\eps$

\be
G_A(e^{i\ph})=1+\lambda\eps N \sum_{k=1}^N \cos \ph_k +0(\eps^2)
\label{pot}
\ee
in order to have a proper continuum limit.
In \eq{pot} we have neglected all representations with more than one
box in the Young table.  It means that we allow only for the 1-fold
coverings of the disks. To do it, we have to tend the areas of the
disks, $A$, to the infinity as: $e^{-A}\equiv\lambda\eps\ll1/N$.

In the continuum limit, we find the fermionic path integral

\be
{\cal Z}=\int \prod_{k=1}^N {\cal D}\ph_k\;
\exp \int dt\; N\sum_{k=1}^N\bigg(-\frac12\stackrel{\cdot}{\ph}^2_k
+\lambda \cos \ph_k\bigg)
\ee

This problem is equivalent to solving the following Schr\"{o}dinger
equation

\be
\frac{\d^2\ }{\d\ph^2}\psi_n(\ph)+2N^2(e_n+\lambda\cos\ph)\psi_n(\ph)=0
\label{schrodeq}
\ee
with the periodic boundary conditions $\psi_n(\ph+2\pi)=\psi_n(\ph)$.

The $N$-fermion wave function is given by the Slater determinant

\be
\Psi(\ph_1,\ldots,\ph_\N)=\frac1{N!}\det[\psi_i(\ph_j)]
\ee
with the ground state energy equal to the sum of $N$ lowest levels:

\be
E=N\sum_{n=1}^N e_n
\ee

The finite temperature free energy is simply

\be
F(\mu,\beta)=\sum_{n=1}^{\infty}\log(1+e^{\N(\mu-\beta e_n)})
\ee
where $N\mu$ is a chemical potential and $\beta$ is an inverse
temperature.

The spectrum of \eq{schrodeq} is discrete and the one-particle
stationary states are described by the Mathieu wave functions.
However, we are interested only in the large $N$ limit, where the
stringy interpretation exists. Therefore, we can simplify considerably
the problem by treating \eq{schrodeq} quasiclassicly.

The large $N$ wave functions are

\be
\psi_n(\ph)\approx \mbox{$\frac{{\displaystyle e^{iNS(\ph)}}}
{\sqrt{S'(\ph)}}$}
\ee
where the phase is given by

\be
S=\int d\ph \sqrt{2(e+\lambda\cos\ph)}
\ee
and the classical dynamics is described by the equation

\be
t=\int\frac{d\ph}{\sqrt{2(e+\lambda\cos\ph)}}
\label{t=}
\ee
which solves in elliptic functions. Let us introduce the new variable
$x=\sin\frac{\ph}2$. Then \eq{t=} takes the form

\be
\mbox{$\sqrt{\frac{e+\lambda}2}$}\; t=
\int\frac{dx}{\sqrt{(1-x^2)(1-\frac{2\lambda}{e+\lambda}x^2)}}
\ee
whose solution is the elliptic sinus:
$x={\rm sn}(\sqrt{\frac{e+\lambda}2}t)$
with the modulus $k=\sqrt{\frac{2\lambda}{e+\lambda}}$.

The quasiclassical quantization gives the equation

\be
{\rm Re}\int_{-\pi}^{+\pi} d\ph \sqrt{2(e_n+\lambda)-4\lambda\sin^2
\frac{\ph}2}=\left\{\ba{ll}
\pi\frac{n+\frac12}N,\ & |e_n|<\lambda\\
2\pi\frac{n}N,\ & e_n>\lambda
\ea\right.
\ee
The levels $e_n>\lambda$ are twice degenerate.

Let us rescale the energy $e\to\lambda e$ and introduce the density of
energy levels

\be
\rho(e)=\frac1{\sqrt{\lambda}N} \frac{\d n}{\d e}=
\mbox{$\frac2{\pi}\sqrt{\frac2{e+1}}$}{\rm Re}\;
K\Big(\mbox{$\sqrt{\frac2{e+1}}$}\Big)
\ee
where $K(k)$ is the complete elliptic integral of the first kind. This
expression is valid for all values of $e$ and $\lambda$. If the
rescaled energy obeys $|e|<1$, it is convenient to introduce the
inverse modulus: $\wt{k}=\frac1k=\sqrt{\frac{e+1}2}$, and then we find

\be
\rho(e)=\left\{\ba{ll}
\frac2{\pi}kK(k), & e>1\\
\frac2{\pi}K(\wt{k}), & |e|<1
\ea\right.
\ee

With the chosen normalization, we find the parametric representation
for the ground state energy ${\cal E}=E/N^2$

\be
\lambda^{-3/2}{\cal E}=\int_{-1}^{e_\F}de\; e\rho(e)
\hspace{2pc}
\lambda^{-1/2}=\int_{-1}^{e_\F}de\; \rho(e)
\ee
where $e_\F$ is a Fermi level. We are looking for a singularity of
${\cal E}(\lambda)$. It is convenient to differentiate
$\lambda^{-3/2}{\cal E}$ twice with respect to $\lambda^{-1/2}$
\cite{KM}, then

\be
\frac{\d^2 \lambda^{-3/2}{\cal E}}{(\d \lambda^{-1/2})^2}=\frac1{\rho(e_\F)}
\ee

The critical point corresponds to $e_\F=1$. Using the standard
formulas we find the asymptotics

\be
\rho(e)=\left\{\ba{ll}
\frac1{\pi}\log\frac{32}{1-e}+\frac1{4\pi}(\log\frac{32}{1-e}-2)(1-e)
+O\Big((1-e)^2\Big),&\ 0<1-e\ll1\\
\frac1{\pi}\log\frac{32}{e-1}+O\Big((e-1)^2\Big), &\ 0<e-1\ll1
\ea\right.
\ee
Thus we find the third order phase transition.

\section{Discussion}

The only universal feature of the model considered in the previous
section is the $c=1$ string phase transition, which takes place when
the Fermi level reaches the maximum of the potential \cite{KM}. Its
existence and the type of the singularity do not depend upon a choice
of the Boltzmann weight (any periodic function has a maximum).
Therefore, whatever a lattice action would be, it produces in the
continuum limit the non-critical Bose string. We associate this
universal stringy behavior with a glueball threshold.  If multicolor
QCD can indeed be reformulated as a string model, the lowest glueball
has to represent a closed-string state.

At a particle threshold, the classical dynamics of a field theory
becomes effectively 1-dimensional, simply because the energy is of the
order of a mass. Of course, it is not true quantum mechanicly.
However, in string theory, quantum loop effects are associated with
higher topologies. Therefore, we expect that, at the tree level, the
one-dimensional string dynamics could correctly describe some universal
features of higher-dimensional gauge models.

Another argument in favor of this conclusion is provided by the
Fateev-Kazakov-Wiegmann exact solution of Principal Chiral Field at
large $N$ \cite{FKW}. They have found that the threshold singularity of the
PCF free energy is identical to the one in the $c=1$ matrix model. To
make the parallel between PCF and the model considered in the present
paper more transparent, let us consider the infinite 2-dimensional
lattice of cubes. One can imagine it as two parallel square lattices
whose vertices are connected pairwise by ``vertical'' links. Let us
consider the standard gauge theory on this ``two-layer'' lattice. To
find a continuum limit, we have to introduce different coupling
constants for ``vertical'' and ``horizontal'' plaquettes, $g_v$ and
$g_h$ respectively. If $g_v=\infty$, the model is equivalent to 2
non-interacting copies of QCD$_2$. If $g_h=0$, we find a lattice
regularization of PCF. Therefore, in the continuum limit, we find the
action

\be
{\cal A}=\int d^2x\;
N\tr\Big(\frac1{2g_v}(\nabla^{\scs{A}}_{\alpha}\phi)
(\overline{\nabla}^{\scs{B}}_{\alpha}\phi^{-1})+
\frac1{2g_h}(F^{\scs{A}}_{\mu\nu})^2+
\frac1{2g_h}(F^{\scs{B}}_{\mu\nu})^2\Big)
\label{coma}
\ee
where $\phi(x)\in SU(N)$ is PCF;
$\nabla^{\scs{A}}_{\mu}=\d_{\mu}+iA_{\mu}$
($\nabla^{\scs{B}}_{\mu}=\d_{\mu}+iB_{\mu}$) are the covariant
derivatives and
$F^{\scs{A}}_{\mu\nu}=[\nabla^{\scs{A}}_{\mu},\nabla^{\scs{A}}_{\nu}]$
($F^{\scs{B}}_{\mu\nu}=[\nabla^{\scs{B}}_{\mu},\nabla^{\scs{B}}_{\nu}]$)
are the curvature tensors for 2 copies of gauge field, $A_{\mu}$ and
$B_{\mu}$, respectively.

The model simplifies in the axial gauge $A_1=B_2=0$. After integrating
out the gauge fields, one finds the effective action for $\phi(x)$

\be
{\cal A}^{ef\! f}=\int d^2x\;
N\tr\Big\{\frac1{2g_v}|\d_{\alpha}\phi|^2 +\frac{g_h}{2g_v^2}
\Big(J_1^L\frac1{\d^2_2}J_1^L + J_2^R\frac1{\d_1^2}J_2^R\Big) \Big\}
\ee
where $J_{\alpha}^L=\phi^{-1}\d_{\alpha}\phi$ and
$J_{\alpha}^R=\d_{\alpha}\phi\phi^{-1}$ are the left and right
invariant currents. This model possesses the main qualitative features
of QCD: asymptotic freedom, confinement and non-trivial glueball
spectrum.  Unfortunately, it looks too complicated to be solved
exactly. Fateev, Kazakov and Wiegmann have investigated PCF in the
homogeneous external gauge field which fixes an energy scale. It is
very plausible that, at the threshold, the model (\ref{coma}) behaves
identically to this simplified one. If we accept this hypothesis, it
will presumably mean that, despite the sum-over-surfaces formulation
of lattice gauge theory looks highly non-trivial \cite{Kaz,D}, what
shows in the continuum limit of multicolor QCD is the
simplest non-critical string. However, in higher dimensions, it
becomes tachionic thus disappearing and leaving us to guess what
really happens in physics.

\bigskip

{\Large \bf Acknowledgments}
\bigskip\nopagebreak

I thank B.E.Rusakov for the numerous fruitful discussions. This work was
supported by the EEC program ``Human Capital and Mobility'' under the
contract ERBCHBICT9941621.

\end{document}